\documentclass[letterpaper,10pt]{article} 
\usepackage{opticameet3} %% use version 3 for proper copyright statement

\newcommand\authormark[1]{\textsuperscript{#1}}

\usepackage{amsmath,amssymb}
\usepackage[colorlinks=true,bookmarks=false,citecolor=blue,urlcolor=blue]{hyperref} %pdflatex

\begin{document}

\title{Optimization of continuous-wave NIRS devices for placental monitoring. A simulation study}

\author{C. Caredda,\authormark{1,*} F Lange,\authormark{2} U Hakim,\authormark{2} N. Ranaei-Zamani,\authormark{3} A. L David,\authormark{3} D. Siassakos,\authormark{3} R. Aughwane,\authormark{3} S. Hillman,\authormark{3} K. Olayinka,\authormark{3} S. Mitra\authormark{3} and I. Tachtsidis\authormark{2}}

\address{\authormark{1} Univ Lyon, INSA-Lyon, Universit\'e Claude Bernard Lyon 1, UJM-Saint Etienne, CNRS, Inserm, CREATIS UMR 5220, U1294, F69100, Lyon, France\\
	\authormark{2}Department of Medical Physics and Biomedical Engineering, University College London, UK\\
	\authormark{3}EGA Institute for Women's Health, University College London, London, United Kingdom}

\email{\authormark{*}charly.caredda@creatis.insa-lyon.fr} %% email address is required

\begin{abstract}
	Near infrared spectroscopy (NIRS) is an optical technique that is widely used to monitor tissue oxygenation. These devices take advantage of the near infrared light to monitor deep tissues like brain, muscle or placenta. In this study, we developed a Monte Carlo framework to evaluate the sensitivity of continuous-wave (CW) NIRS devices for monitoring the placenta which a deep layer in the maternal abdomen. This framework can be used to optimize CW-NIRS acquisition parameters (integration time, source detector separation) before going into clinical applications.
\end{abstract}

%\keywords{Manuscript format, template, SPIE Proceedings, LaTeX}

\section{Introduction} 

Abnormal placental development is a major cause of adverse pregnancy outcomes, such as hypertensive disorders, fetal growth restriction, and stillbirth \cite{placenta_clinics2}. These adverse outcomes are associated with poor placental oxygen perfusion \cite{placenta_clinics}, which is challenging to assess during clinical practice.\\

Near-infrared spectroscopy (NIRS) is an optical technique widely used to monitor tissue oxygenation. These devices take advantage of near-infrared light to monitor deep tissues such as the brain, muscles, and placenta \cite{CYRIL}. NIRS can measure oxy-hemoglobin ($HbO_2$) and deoxy-hemoglobin ($Hb$) concentrations and has been extensively employed to monitor tissue hemodynamics \cite{fNIRS_review1}. Frequency-domain NIRS has been used to monitor placental oxygen saturation during maternal gestation. The NIRS device was integrated with ultrasound imaging to model light propagation in a multi-layered medium in order to quantify hemodynamics in the placenta layer. Industrial systems, such as CW-NIRS devices, can also monitor placental blood oxygenation \cite{CW_NIRS}. For the calculation of the placental oxygenation, the spatially resolved spectroscopy (SRS) technique \cite{SRS} is used, which requires assumptions about tissue homogeneity. Since the maternal abdomen comprises layered structures, the estimation of absolute concentration of $HbO_2$ and $Hb$ by the SRS algorithm is less accurate. \cite{fNIRS_review3}. The light collected by CW devices includes contributions from different layers up to approximately 2 cm in depth (skin, adipose tissue, muscle, and placenta), but the proportions of these contributions cannot be determined from the measurements.\\

In developing NIRS technologies for placental monitoring, tools and methodologies are required to investigate photon sensitivity and measurement accuracy within the multilayered abdomen. To better understand the possibilities and current limitations of CW-NIRS devices, we developed a Monte Carlo framework to evaluate the sensitivity of CW-NIRS devices for monitoring placental function. This framework can optimize acquisition parameters (e.g., integration time, source-detector separation) before clinical application and identify study cases (e.g., tissue thickness, skin tone) suitable for monitoring placental function.

 \section{Material and methods}
	
	The Monte Carlo pipeline for evaluating the sensitivity of CW-NIRS devices in monitoring placental function is divided into two parts: (1) Ultrasound images were used for measuring tissue thickness and placenta depth. Monte Carlo simulations were used to model the light propagation in the maternal abdomen and the acquisition of the retro-diffused light with a multi-distance broadband NIRS device (Mini CYRIL \cite{Mini_CYRIL}). With these simulations, the sensitivity probability was calculated for each tissue layer. (2) We calculated the probability of a detector of the CW-NIRS device to acquire a signal that is significantly above the noise level (detection probability). This probability was calculated with a calibration procedure with a solid phantom.

	\subsection{Model of maternal abdomen}
\begin{figure}[!h]
		\centering
		\begin{tabular}{ccc}
			A & B & C \\
			\includegraphics[height=4cm]{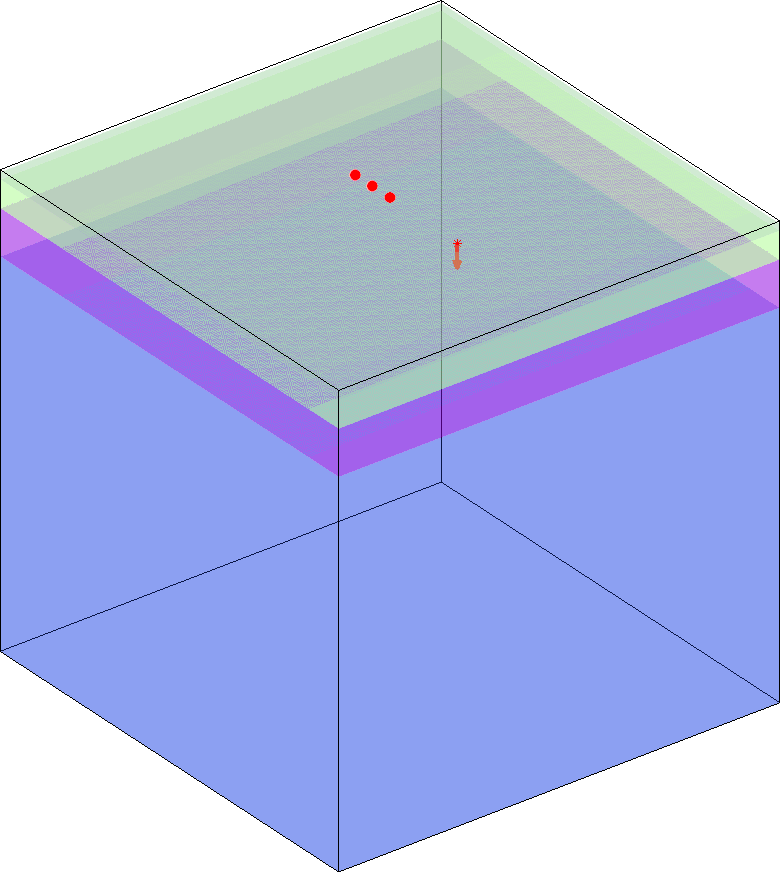} &
			\includegraphics[height=4cm]{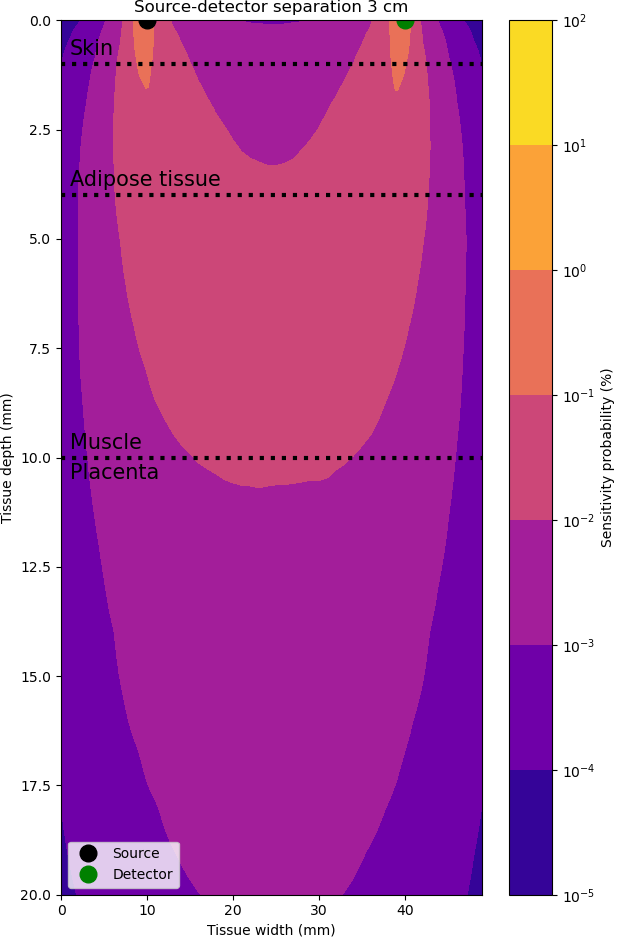} &
			\includegraphics[height=4cm]{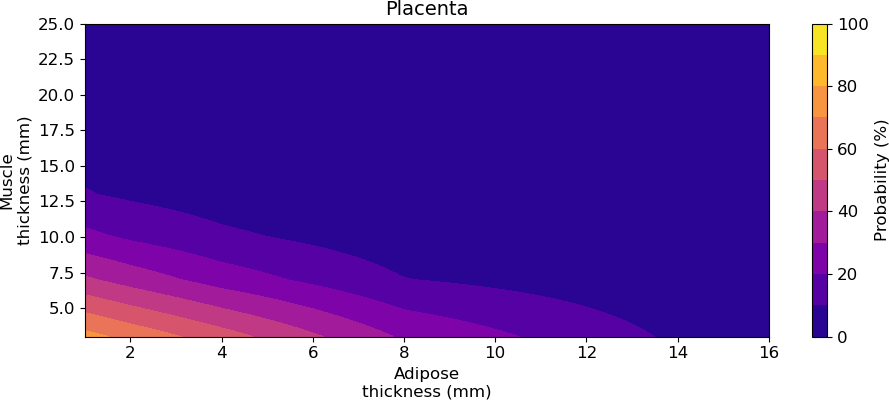}
		\end{tabular}
		
		\caption{Monte Carlo pipeline for optimizing NIRS measurements for placenta monitoring. A - Monte Carlo model representing the tissue and the NIRS device (skin: dark green, adipose tissue: light green, muscle: purple, placenta: blue). B - Sensitivity profile for a source-detector separation of 3 cm. C - Placenta sensitivity (probability to scan the placenta in $\%$) in function of tissue thickness for a source-detector separation of 3 cm.}
		\label{fig}
	\end{figure}

	We modelled the maternal abdomen as a 4-layers volume (skin, adipose tissue, muscle, and placenta). The volume dimensions were $200 \times 200 \times 200$ mm with a resolution of $1$ mm$^{3}$, as shown in Fig.~\ref{fig} A. Each voxel of the modelled tissue contained information about its optical properties: absorption coefficient ($\mu_a$ in $cm^{-1}$), reduced scattering coefficient ($\mu_s'$ in $cm^{-1}$), anisotropy coefficient ($g$), and refractive index ($n$). Using the MCX software \cite{MCX}, we simulated the propagation of light emitted at $780$ nm in the maternal abdomen and its detection with a multi-distance NIRS device (Mini CYRIL \cite{Mini_CYRIL}). A point light source was positioned perpendicular to the surface of the volume, emitting $10^9$ packets of photon. Eight detectors, spaced at distances ranging from $3$ cm to $5$ cm from the source, were created. These detectors collected the retro-diffused light within a disk area of $1$ mm in diameter. A total of $51840$ Monte Carlo simulations were performed with varying parameters (layer thickness, melanin volume fraction, muscle and placenta blood volume and oxygen saturation). Measurements of skin tones, and thickness of the skin, adipose tissue and muscle layers were performed on $268$ healthy participants during maternal gestation at the Institute for Women's Health in the University College London Hospital (UK). The measurements were approved by the local ethics committee of University College London Hospital and the participating patients signed written consent.

	\subsection{Sensitivity and detection probabilities}
	We used the fluence rate and the adjoint Method \cite{sensitivity_profiles_MC} to calculate sensitivity matrices. By normalizing the sensitivity matrix by its sum, we obtained the probability volume $P$ of photon propagation between a pair of point-source and point-detector for a perfect device (infinite $SNR$). The probability of the detector $D$ to scan the tissue $N$ was calculated by summing the probability volume $P$ on the slice of the tissue $N$:
	
	\begin{equation}
		S^D_N = \sum_x \sum_y \sum_{z={z_{N_1}}}^{z_{N_2}} P(x,y,z),
		\label{Eq:sensitivity_indexes}
	\end{equation}

	\noindent with $z_{N_1}$ and $z_{N_2}$ the depth indexes of the tissue $N$. We defined the detection probability as the probability of a detector to acquire a signal that is significantly above the noise level. This probability was calculated with a calibration procedure:

	\noindent \textbf{1)} We performed diffuse reflectance measurements the CW-NIRS device Mini CYRIL at $780$ nm of a solid multi-layered phantom. The multi-layered phantom was composed of three solid layers and a bloc with known optical properties matching those of the maternal abdomen.

	\noindent \textbf{2)} We performed Monte Carlo simulations of the multi-layered phantom to obtain the simulated diffuse reflectance at $780$ nm for each detector.

	\noindent \textbf{3)} A normalization step was performed to scale measured quantities to simulated diffuse reflectance
		
	\noindent \textbf{4)} The minimum detectable diffuse reflectance $\phi_{min}$ was obtained for each detector with the standard deviation of the scaled measured quantities.
	
	\noindent \textbf{4)} The sensor noise was modelled as a normal distribution with a mean of $0$ and a standard deviation of $\sqrt{\phi_{min}}$ \cite{standard_camera}.
	
	\noindent \textbf{5)} The detection probability was obtained with a one-sided T-test.\\
	
	Finally, we defined the scanning probability as the probability of detector $D$ scanning tissue $N$. This was calculated by multiplying the detection probability by the sensitivity probability $S^D_N$ (independent probabilities). It represents the probability that detector $D$ receives the minimum detectable irradiation and that tissue $N$ is scanned for the given source-detector separation.
	
	\section{Results and discussion}
	
	CW-NIRS device can be used to monitor the placenta but precautions must be taken to avoid ambiguities in the analysed signal. Ultrasound imaging is necessary before NIRS acquisition to evaluate placental scanning probability. Muscle layer significantly impacts the collected signal, potentially leading to erroneous conclusions. CW-NIRS devices are more effective for placentas closer to the surface. The Monte Carlo framework takes into account the skin tone of the participants. We observed that skin tones affect photon depth penetration and signal to noise ratio (detection probability). Feiner et al. showed that a dark skin tone induce a decrease in the accuracy of pulse oximeters for measuring oxygen saturation \cite{skin_effect3}. This Monte Carlo framework may be use to identify these measurement biases.

	\section{Conclusion} This Monte Carlo framework is a powerful tool to evaluate the capability of NIRS device to monitor deep tissue oxygenation. We showed that a source-detector separation of 3 or 4 cm can be used to monitor the placental function under certain conditions (layer thickness, and blood volume). This tool will be improved in the future to reduce calculation time ($\approx$ 1 week on a computation grid using a NVIDIA GK210 GPU).

\section*{Acknowledgments} % equivalent to \section*{ACKNOWLEDGMENTS}       
This work was supported by Wellcome LEAP; Labex Primes (ANR-11-LABX-0063/ANR-11-IDEX-0007); Infrastructures d’Avenir en Biologie Santé (ANR-11-INBS-000); France Life Imaging (ANR-11-INBS-0006).

% References
\bibliographystyle{unsrt} % We choose the "plain" reference style
\bibliography{report}

\end{document}